\documentclass[prd,twocolumn,showpacs,preprintnumbers,amsmath,amssymb,superscriptaddress,nofootinbib,english]{revtex4-2}
\usepackage[utf8]{inputenc}
\usepackage{comment}
\usepackage{soul}
\usepackage{graphicx}
\usepackage{longtable}
\usepackage{amsmath}
\usepackage{color} 
\usepackage{amssymb}
\usepackage{subfigure}
\usepackage{tabularx}
\usepackage{microtype}
\usepackage[linktoc=all]{hyperref}
\hypersetup{
    colorlinks=true,
    linkcolor=BlueViolet,
    citecolor=Blue,
    urlcolor=MidnightBlue,
}
\usepackage{cleveref}
\usepackage{stackengine}
\usepackage[normalem]{ulem}
\usepackage{slashed}
\usepackage{color}
\usepackage[dvipsnames]{xcolor}

\usepackage{multirow}
\usepackage{array}
\usepackage{placeins}
\usepackage{dcolumn}
\usepackage{subfloat}

\definecolor{greenW}{rgb}{0.0, 0.55, 0.1} 

\begin{document}

\title{Cosmological Impact of Redshift-Dependent Type Ia Supernovae Calibration}

\author{Seyed Hamidreza Mirpoorian}
\email{smirpoor@sfu.ca}
\affiliation{Department of Physics, Simon Fraser University, Burnaby, British Columbia, V5A 1S6, Canada}
\affiliation{Laboratoire de Univers et Particules de Montpellier, UMR5299-CNRS, Universite de Montpellier, 34095 Montpellier, France}

\author{Meng-Xiang Lin}
\email{mengxiang\_lin@sfu.ca}
\altaffiliation{CITA National Fellow}
\affiliation{Department of Physics, Simon Fraser University, Burnaby, British Columbia, V5A 1S6, Canada}
\affiliation{Canadian Institute for Theoretical Astrophysics (CITA), University of Toronto, 60 St George Street, Toronto, Ontario M5S 3H8, Canada}

\author{Levon Pogosian}
\email{levon\_pogosian@sfu.ca}
\affiliation{Department of Physics, Simon Fraser University, Burnaby, British Columbia, V5A 1S6, Canada}


\begin{abstract}
Type Ia supernovae (SNIa) play a central role in constraining the late-time expansion history of the Universe and are directly implicated in current cosmological tensions. Motivated by the possibility of unaccounted redshift-dependent calibration systematics or new physics, we investigate the impact of a phenomenological correction to SNIa magnitudes that scales with cosmic look-back time. We parameterize this effect with a free amplitude and constrain it using a combination of cosmic microwave background, baryon acoustic oscillation, and SNIa data, considering both $\Lambda$CDM and dynamical dark energy models. Importantly, our parameterization is not intended to serve as a proxy for SNIa progenitor age, as current observations show no significant difference in standardized SNIa brightness between young and old progenitor populations at low redshift. We find no evidence for a redshift-dependent calibration effect when fitting uncalibrated SNIa data, and its inclusion has a negligible impact on cosmological parameters within $\Lambda$CDM, nor does it qualitatively change the inferred dynamics of evolving dark energy. When incorporating a prior on the SNIa absolute magnitude from SH0ES, a nonzero calibration parameter is weakly preferred within $\Lambda$CDM. Interestingly, with dynamical dark energy, the preference of a nonzero calibration parameter increases to $4.3\sigma$, and it can accommodate both the distance ladder and early-Universe constraints, reducing the Hubble tension to $1.5\sigma$, with the best-fit model effectively corresponding to a constant equation of state with $w < -1$. Overall, our results indicate that redshift-dependent SNIa calibration effects, as parameterized here, are not supported by current data within $\Lambda$CDM, but can play a role in reconciling cosmological datasets when combined with extensions to the late-time expansion history.
\end{abstract}
\maketitle

\section{Introduction}

Recent cosmological observations continue to show evidence of deviations from the standard $\Lambda$ Cold Dark Matter ($\Lambda$CDM) paradigm~\cite{Leauthaud:2025azz}. In particular, the Hubble tension, a discrepancy in the value of the Hubble constant inferred from the Cosmic Microwave Background (CMB)~\cite{Planck:2018vyg,SPT-3G:2025bzu} and the one measured from distance-ladder-calibrated Type Ia supernovae (SNIa)~\cite{Riess:2021jrx,Breuval:2024lsv}, and evidence for dynamical dark energy in the Baryon Acoustic Oscillations (BAO) data from the Dark Energy Spectroscopic Instrument (DESI)~\cite{DESI:2025zgx}, have attracted significant attention. Many ideas for new physics beyond $\Lambda$CDM were proposed to address these cosmological tensions, see, e.g.~\cite{Kamionkowski:2022pkx,CosmoVerseNetwork:2025alb,Cai:2025mas,Arza:2026rsl} and references there. 

SNIa datasets play a key role in both the Hubble tension and the evidence for dynamical dark energy. The distance measurements using SNIa rely on empirically established relations between their peak brightness and the width of the light curve~\cite{Phillips:1993ng} and color~\cite{Riess:1996pa,Tripp:1997wt}, assuming that the empirical relations established from low-redshift observations also apply at high redshifts. The standardization of SNIa has been subject to tremendous levels of scrutiny over multiple decades, and modern analyses account for a host of possible environmental effects (see \cite{Wiseman:2026nty,Murakami:2026knz} and references therein). Within the standard cosmological framework, these studies find no evidence for residual biases once known systematics are included. In this work we instead take a phenomenological approach and ask whether new physics beyond the standard framework could give rise to an effective redshift-dependent modulation of SNIa magnitudes that would impact cosmological interpretations. For example, the intrinsic luminosity of SNIa may vary if the effective Newton’s constant evolves with redshift, as in certain modified gravity scenarios~\cite{Amendola:1999vu,Wright:2017rsu,Ruchika:2023ugh}, or through a screened fifth force mediated by dark matter–baryon interactions~\cite{Sakstein:2019qgn}.

The potential impact of redshift-dependent effects on supernovae cosmology has long been recognized. Early studies showed that phenomenological evolution of the standardized SNIa luminosity can be strongly degenerate with the inferred cosmological parameters~\cite{Drell:1999dx}, while nearly gray intergalactic dust was considered as a possible explanation for the dimming of SNIa without invoking cosmic acceleration~\cite{Aguirre:1999dc}. More general analyses showed that, when luminosity evolution is allowed, a non-accelerating Universe can fit SNIa, BAO, and CMB observations comparably well to the $\Lambda$CDM model~\cite{Tutusaus:2018ulu, Tutusaus:2017ibk}. Although these studies do not establish the presence of an unmodeled bias in SNIa samples, they illustrate the sensitivity of cosmological inference to any redshift-dependent shift in SNIa magnitudes and motivate constraining such an effect jointly with cosmology.

Several complementary studies have examined redshift-dependent SNIa effects using different approaches~\cite{Dhawan:2024gqy}. An analysis of the Union2.1 sample reported a redshift dependence of the color-luminosity relation and demonstrated its impact on the inferred cosmological parameters~\cite{Mohlabeng:2013gda}. Another study parameterized intrinsic-magnitude evolution as $\Delta m(z)=\epsilon z^\delta$ and showed that it can be strongly degenerate with $H_0$ and dark-energy parameters, although these degeneracies are substantially reduced and constraints are considerably improved when BAO, cosmic chronometer, and CMB lensing data are included~\cite{DiValentino:2020evt}. Model-independent reconstructions of intrinsic magnitude $M_b(z)$ using SNIa and BAO measurements found only a weak indication of redshift evolution and no decisive preference over a constant $M_b$ in another parametric analysis~\cite{Benisty:2022psx}. 
A redshift-dependent correction to SNIa magnitude can also be interpreted as an effective violation of the distance duality relation (DDR, the relation between luminosity distance and angular diameter distance)~\cite{Tutusaus:2023cms,Teixeira:2025czm, Dhawan:2025mer}.
More recently, a consistency test based on the DDR reported a redshift-dependent discrepancy between Pantheon+ and DESI DR2 and showed that accounting for this effect shifts the inferred dark energy parameters toward $\Lambda$CDM~\cite{Afroz:2025iwo}. From the perspective of possible BAO systematics, a redshift-independent rescaling of the BAO calibration chosen to shift the inferred $H_0$ towards the SH0ES value was found to be unable to accommodate late-time solutions to the Hubble tension once uncalibrated SNIa and CMB geometrical constraints were included~\cite{Pedrotti:2025ccw}.

The impact of a redshift-dependent correction to SNIa magnitude on the evidence for dynamical dark energy from DESI BAO was investigated recently in~\cite{Son:2025rdz}, motivated by a bias correlated to the SNIa progenitor age with a fixed amplitude. However, subsequent analyses~\cite{Wiseman:2026nty,Murakami:2026knz,Popovic:2026doa} pointed out that there is no evidence for such an ``age bias'' in SNIa datasets after accounting for all systematic effects included in standard analyses. In particular, \cite{Wiseman:2026nty} showed that there is no significant difference at low redshift in standardized brightness between star-forming galaxies and several Gyr older quiescent galaxies of the same stellar mass. This is especially relevant for distance-ladder samples such as SH0ES, which are drawn from low-redshift, star-forming hosts and therefore do not preferentially select old progenitor populations.

In this paper, rather than questioning standard SNIa analyses or proposing new physics that could alter them, we explore the hypothetical impact of a redshift-dependent correction to the SNIa magnitude calibration on the Hubble tension and the evidence for dynamical dark energy. Using a one-parameter phenomenological parameterization, we relate the SNIa magnitude correction to the cosmic look-back time, which is a smooth and converged function of redshift. We fit this model to a combination of CMB, BAO, and SNIa data within the $\Lambda$CDM model and some of its extensions.

We separately consider uncalibrated SNIa data and SNIa data supplemented by the SH0ES prior on the intrinsic magnitude $M_b$. By performing our analysis within both the $\Lambda$CDM and $w_0 w_a$ models, we assess how a possible redshift-dependent correction affects the inferred late-time expansion history, the Hubble tension, and the evidence for the dynamical dark energy.

In Section~\ref{sec:method}, we describe our methods and the cosmological datasets used in the analysis, and present the results in Section~\ref{sec:results}. We conclude with a discussion in Section~\ref{sec:conclusion}.

\section{Methodology and Datasets}\label{sec:method}

To choose a phenomenological parameterization, we want the correction of SNIa magnitude to be a monotonic function of the redshift, zero at today, and finite at $z\rightarrow\infty$ as required by any physical models. The cosmic look-back time is a natural choice for all these requirements, as we show below.
Therefore, we parameterize the redshift-dependent SNIa magnitude correction, $\Delta m(z)$, as
\begin{equation}\label{eq:Deltam}
    \Delta m(z) =  S_{\rm SN}\Delta t(z)\,{\rm mag}/{\rm Gyr}\,,
\end{equation}
where $\Delta t(z)$ is the cosmic look-back time between redshift $z$ and today:
\begin{equation}\label{eq:age}
    \Delta t(z) = \int_0^z \frac{dz'}{(1+z')H(z')}\,,
\end{equation}
and $S_{\rm SN}$ is a dimensionless coefficient. The correction vanishes at $z=0$ by definition, and approaches a finite constant as $z\rightarrow\infty$, as the cosmic time integral is dominated by low redshifts and $\Delta t$ quickly saturates.

Note that $\Delta t(z)$ here is not the SNIa progenitor age as suggested in~\cite{Son:2025rdz}. The calculation of SNIa progenitor age requires knowledge of the cosmic star-formation history and the SNIa explosion delay-time distribution~\cite{Wiseman:2026nty,Murakami:2026knz}, and cosmic look-back time does not provide a realistic description of it. Nevertheless, the parameterization in~\cite{Son:2025rdz} also exhibits a strong redshift dependence, as in our Eq.~\eqref{eq:Deltam}, although with a slightly different shape. Rather than adopting a fixed value of the correction amplitude as in~\cite{Son:2025rdz}, we treat $S_{\rm SN}$ as a free parameter uniformly distributed in $[-1,1]$.

We perform Markov chain Monte Carlo (MCMC) analyses using {\tt Cobaya}~\cite{Torrado:2020dgo} with {\tt CAMB}~\cite{Lewis:1999bs}, considering the $\Lambda$CDM model as well as the $w_0w_a$CDM model, in which the equation of state of dark energy evolves with the scale factor $a$ according to $w(a)=w_0+(1-a)w_a$~\cite{Chevallier:2000qy,Linder:2002et}. 
Other than the bias parameter $S_{\rm SN}$, the dark energy parameters ($w_0,w_a$), and allowing for a non-zero spatial curvature $\Omega_k$, we vary the same standard cosmological parameters as in $\Lambda$CDM, with the base set of six cosmological parameters $(\Omega_b h^2, \Omega_c h^2, H_0, \ln(10^{10}A_s), n_s, \tau)$. We fix the sum of neutrino masses $\sum m_\nu$ to a default value of 0.06 eV. We adopt broad, flat, uninformative priors on all standard $\Lambda$CDM parameters in every analysis in this paper, while using wide flat priors on the parameters $w_0 \in [-3, 1], w_a \in [-3, 2]$ and $\Omega_k \in [-0.3, 0.3]$.
The convergence criterion for MCMC sampling is that the Gelman-Rubin statistic~\cite{Gelman-R} satisfies $R - 1 < 0.02$. Parameter posterios and statistical results are obtained with {\tt getdist} software package~\cite{Lewis:2019xzd}. We determine the maximum a posteriori (MAP) point in parameter space for each model to determine the best-fit points and the corresponding $\chi^2$ values. The MAP for each model is then compared to the best-fit $\Lambda$CDM result for the same dataset combination to calculate $\Delta \chi^2$ values reported in Tables~\ref{t:no_mb} and \ref{t:with_mb}.
We fit the model parameters to the CMB, BAO, and SNIa datasets specified below.

We use a combination of the Planck PR3 low-$\ell$ temperature and polarization spectra~\cite{Planck:2018vyg, Planck:2019nip}, the NPIPE {\tt CamSpec} high-$\ell$ TTTEEE likelihood~\cite{Efstathiou:2019mdh, Rosenberg:2022sdy}, and CMB lensing measurements from a combination of Planck and the Atacama Cosmology Telescope (ACT DR6) data~\cite{Carron:2022eyg, ACT:2023kun, ACT:2023dou}. Throughout the paper, we refer to this dataset combination as PL-AP.  

We use the DESI Data Release 2 BAO data (DESI2)~\cite{DESI:2025zgx} providing 13 standard-ruler distance measurements at 7 effective redshifts, including the bright galaxy survey (BGS)~\cite{Hahn:2022dnf}, luminous red galaxies (LRG)~\cite{DESI:2022gle}, emission line galaxies (ELG)~\cite{Raichoor:2022jab}, and quasars (QSO) with their Lyman-$\alpha$ forests~\cite{Chaussidon:2022pqg}.

We use the Pantheon+ (PP) collection of 1550 Type Ia SN magnitudes spanning $0.001 < z < 2.26$~\cite{Scolnic:2021amr, Brout:2022vxf}, with the added correction given by Eq.~\eqref{eq:Deltam}. We use this dataset without a distance-ladder prior on the intrinsic magnitude in Sec.~\ref{ss:uncalib}, and with the SH0ES intrinsic magnitude prior of $M_b = -19.253 \pm 0.027$~\cite{Riess:2021jrx} in Sec.~\ref{ss:sn_calib}. 
To implement this prior, we use the {\tt Cobaya} implementation of the Pantheon+ SNIa sample with the option {\tt use\_abs\_mag = true}, adopting a Gaussian prior on the intrinsic magnitude, with $M_b \in \mathcal N(-19.253, 0.027)$.

\section{Results}\label{sec:results}
\begin{table*}[htbp!]
        \begin{tabular}{c|cccc}
            \multirow{3}{*}{\bf Name}
            & {\boldmath $\Lambda$}{\bf CDM}
            & {\boldmath $\Lambda$}{\bf CDM}
            & {\boldmath $w_0w_a$}{\bf CDM}
            & {\boldmath $w_0w_a$}{\bf CDM}
            \\
            & PL-AP+DESI2
            & PL-AP+DESI2
            & PL-AP+DESI2
            & PL-AP+DESI2
            \\
            & +PP
            & +PP+bias
            & +PP
            & +PP+bias
            \\
            \colrule
            \rule{0pt}{3ex}
            $H_0$
            & $68.09 \pm 0.27$
            & $68.19 \pm 0.29$
            & $67.54 \pm 0.58$
            & $66.3 \pm 1.5$
            \\
            $100\ \Omega_m h^2$
            & $14.08 \pm 0.06$
            & $14.06 \pm 0.06$
            & $14.20 \pm 0.08$
            & $14.20 \pm 0.08$
            \\
            $\Omega_m$
            & $0.304 \pm 0.004$
            & $0.303 \pm 0.004$
            & $0.311 \pm 0.006$
            & $0.324 \pm 0.015$
            \\
            $S_8$
            & $0.814 \pm 0.008$
            & $0.812 \pm 0.008$
            & $0.827 \pm 0.008$
            & $0.832 \pm 0.010$
            \\
            $w_0$
            & $-$
            & $-$
            & $-0.837 \pm 0.053$
            & $-0.72 \pm 0.13$
            \\[0.5ex]
            $w_a$
            & $-$
            & $-$
            & $-0.63^{+0.22}_{-0.18}$
            & $-0.89^{+0.39}_{-0.33}$
            \\[0.5ex]
            $S_{\rm SN}$
            & $-$
            & $-0.005 \pm 0.003$
            & $-$
            & $0.007 \pm 0.008$
            \\
            $r_d$
            & $147.72 \pm 0.19$
            & $147.75 \pm 0.19$
            & $147.49 \pm 0.21$
            & $147.48 \pm 0.21$
            \\
            \colrule
            \rule{0pt}{3ex}
            $\Delta \chi^2_{\rm CMB}$
            & $-$
            & $+0.34$
            & $-4.22$
            & $-4.68$
            \\[0.5ex]
            $\Delta \chi^2_{\rm BAO}$
            & $-$
            & $-0.41$
            & $-4.10$
            & $-4.06$
            \\[0.5ex]
            $\Delta \chi^2_{\rm SN}$
            & $-$
            & $-2.47$
            & $-1.45$
            & $-0.83$
            \\
            \colrule
            \rule{0pt}{3ex}
            $\chi^2_{\rm total}$
            & 12406.23
            & 12403.69
            & 12396.46
            & 12396.66
            \\
            $\Delta \chi^2_{\rm total}$
            & $-$
            & $-2.54$
            & $-9.77$
            & $-9.57$
        \end{tabular}
    \caption{
    Mean values and 68\% confidence intervals of the cosmological parameters, along with the best-fit $\chi^2$ values, for the $\Lambda$CDM and $w_0w_a$CDM models fitted to the combination of Planck CMB + ACT-DR6 lensing (PL-AP), DESI DR2 BAO (DESI2), and Pantheon+ supernova (PP) data, with and without a redshift-dependent correction to the SNIa magnitude calibration. The $\Delta \chi^2$ values are given relative to the $\Lambda$CDM best fit to PL-AP+DESI2+PP.
    }
    \label{t:no_mb}
\end{table*}

\begin{figure*}[htbp!]
    \includegraphics[width=0.75\textwidth]{figures/cmb_bias2.pdf}
    \caption{
    The 68\% and 95\% confidence contours and the corresponding one-dimensional marginalized posterior distributions for selected parameters in the $\Lambda$CDM and $w_0w_a$CDM models, fitted to combinations of Planck CMB + ACT-DR6 lensing (PL-AP), DESI DR2 BAO (DESI2), Pantheon+ supernova (PP), and PP with the SH0ES intrinsic magnitude prior on $M_b$ (PP+$M_b$), including a redshift-dependent correction to the SNIa magnitude calibration parameterized by $S_{\rm SN}$ (with $S_{\rm SN}=0$ indicated by a dashed line). The vertical bands denote the 68\% and 95\% confidence intervals of the SH0ES prior on $M_b$.
    }
    \label{fig:cmb_bias}
\end{figure*}

\begin{table*}[!ht]
        \begin{tabular}{c|cccccc}
            \multirow{3}{*}{\bf Name}
            & {\boldmath $\Lambda$}{\bf CDM} 
            & {\boldmath $\Lambda$}{\bf CDM}
            & {\boldmath $\Lambda${\bf CDM}+$\Omega_k$}
            & {\boldmath $w_0w_a$}{\bf CDM}
            & {\boldmath $w_0w_a$}{\bf CDM}{\boldmath $+\Omega_k$}
            & {\boldmath $w$}{\bf CDM}
            \\
            & PL-AP+DESI2
            & PL-AP+DESI2
            & PL-AP+DESI2
            & PL-AP+DESI2
            & PL-AP+DESI2
            & PL-AP+DESI2
            \\
            & +PP+$M_b$
            & +PP+$M_b$+bias
            & +PP+$M_b$+bias
            & +PP+$M_b$+bias
            & +PP+$M_b$+bias
            & +PP+$M_b$+bias
            \\
            \colrule
            \rule{0pt}{3ex}
            $H_0$
            & $68.53 \pm 0.27$
            & $68.65 \pm 0.27$
            & $69.04 \pm 0.31$
            & $71.30 \pm 0.85$
            & $71.40 \pm 0.87$
            & $71.42 \pm 0.68$
            \\
            $100\ \Omega_m h^2$
            & $14.01 \pm 0.06$
            & $13.99 \pm 0.06$
            & $14.24 \pm 0.11$
            & $14.20 \pm 0.08$
            & $14.29 \pm 0.11$
            & $14.19 \pm 0.07$
            \\
            $\Omega_m$
            & $0.298 \pm 0.004$
            & $0.297 \pm 0.003$
            & $0.299 \pm 0.004$
            & $0.279 \pm 0.007$
            & $0.280 \pm 0.007$
            & $0.278 \pm 0.005$
            \\
            $\Omega_k$
            & $-$
            & $-$
            & $0.0032 \pm 0.0012$
            & $-$
            & $0.0016 \pm 0.0012$
            & $-$
            \\
            $S_8$
            & $0.806 \pm 0.007$
            & $0.804 \pm 0.007$
            & $0.814 \pm 0.008$
            & $0.813 \pm 0.009$
            & $0.815 \pm 0.009$
            & $0.812 \pm 0.007$
            \\
            $w_0$
            & $-$
            & $-$
            & $-$
            & $-1.106 \pm 0.077$
            & $-1.126 \pm 0.083$
            & $-1.125 \pm 0.028$
            \\[0.5ex]
            $w_a$
            & $-$
            & $-$
            & $-$
            & $-0.07^{+0.26}_{-0.23}$
            & $0.05^{+0.29}_{-0.25}$
            & $-$
            \\[0.5ex]
            $S_{\rm SN}$
            & $-$
            & $-0.009 \pm 0.003$
            & $-0.008 \pm 0.003$
            & $-0.017 \pm 0.004$
            & $-0.017 \pm 0.004$
            & $-0.018 \pm 0.003$
            \\
            $r_d$
            & $147.81 \pm 0.19$
            & $147.86 \pm 0.19$
            & $147.36 \pm 0.26$
            & $147.47 \pm 0.21$
            & $147.28 \pm 0.26$
            & $147.49 \pm 0.19$
            \\
            $M_b$
            & $-19.405 \pm 0.008$
            & $-19.382 \pm 0.011$
            & $-19.371 \pm 0.012$
            & $-19.308 \pm 0.023$
            & $-19.306 \pm 0.023$
            & $-19.305 \pm 0.020$
            \\
            \colrule
            \rule{0pt}{3ex}
            $\Delta \chi^2_{\rm CMB}$
            & $-$
            & $+2.88$
            & $-4.77$
            & $-7.72$
            & $-5.99$
            & $-7.67$
            \\[0.5ex]
            $\Delta \chi^2_{\rm BAO}$
            & $-$
            & $-0.42$
            & $+1.42$
            & $+8.85$
            & $+6.08$
            & $+6.95$
            \\[0.5ex]
            $\Delta \chi^2_{\rm SN}$
            & $-$
            & $-2.24$
            & $-1.95$
            & $-1.68$
            & $-1.81$
            & $-3.01$
            \\
            $\Delta \chi^2_{\rm SH0ES}$
            & $-$
            & $-9.43$
            & $-13.13$
            & $-28.40$
            & $-28.71$
            & $-26.08$
            \\
            \colrule
            \rule{0pt}{3ex}
            $\chi^2_{\rm total}$
            & 12440.28
            & 12431.07
            & 12421.85
            & 12411.33
            & 12409.85
            & 12410.47
            \\
            $\Delta \chi^2_{\rm total}$
            & $-$
            & $-9.21$
            & $-18.43$
            & $-28.95$
            & $-30.43$
            & $-29.81$
        \end{tabular}
    \caption{
    Mean values and 68\% confidence intervals of the cosmological parameters, along with the best-fit $\chi^2$ values, for the $\Lambda$CDM, $\Lambda$CDM+$\Omega_k$, $w_0w_a$CDM, and $w_0w_a$CDM+$\Omega_k$ models fitted to the combination of Planck CMB + ACT-DR6 lensing (PL-AP), DESI DR2 BAO (DESI2), and Pantheon+ supernova data with the SH0ES intrinsic magnitude prior on $M_b$ (PP+$M_b$), allowing for a redshift-dependent correction to the SNIa magnitude calibration. The $\Delta \chi^2$ values are given relative to the $\Lambda$CDM best fit to PL-AP+DESI2+PP+$M_b$.
    }
    \label{t:with_mb}
\end{table*}

\begin{figure}[htpb!]
    \centering
    \includegraphics[width=0.45\textwidth]{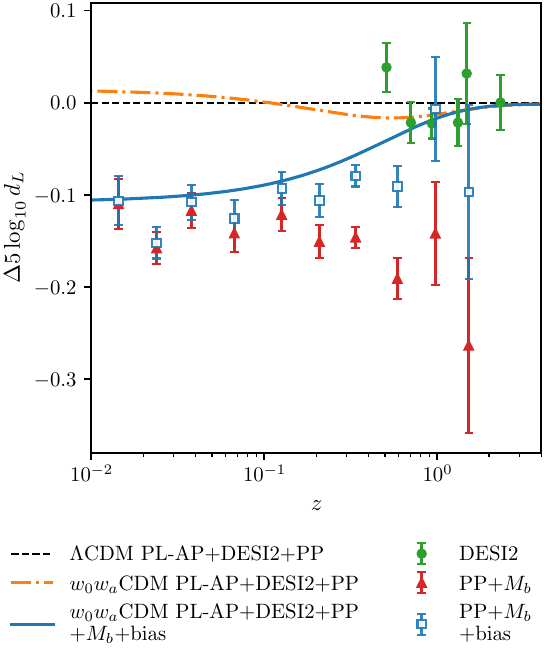}
    \caption{
    The luminosity distances predicted by the $w_0w_a$CDM best fit to the combination of PL-AP, DESI2, PP, and $M_b$, allowing for a redshift-dependent correction to the SNIa magnitude calibration (blue solid), compared to the $\Lambda$CDM (black dashed) and $w_0w_a$CDM (orange dash-dot) best fits to PL-AP+DESI2+PP without the correction. Also shown are the transverse DESI DR2 BAO data points calibrated using the standard ruler from the baseline $\Lambda$CDM PL-AP+DESI2+PP model (green circles), Pantheon+ data points calibrated with the SH0ES intrinsic magnitude prior on $M_b$ (red triangles), and the SNIa data points anchored to the SH0ES measurement of $M_b$ using the best-fit correction with $S_{\rm SN} = -0.018$ in the $w_0w_a$CDM model (blue squares).
    }
    \label{fig:lumdist_combine}
\end{figure}

In what follows, we investigate the impact of allowing for the redshift-dependent SNIa magnitude bias on cosmological parameters. To assess the implications for dynamical dark energy and the Hubble tension, we separately consider the cases with and without including the SH0ES measurement of $M_b$.

\subsection{Cosmology with redshift-dependent bias and uncalibrated supernova}\label{ss:uncalib}

We fit the $\Lambda$CDM and the $w_0w_a$CDM models to the combination of CMB, BAO and uncalibrated PP SNIa data, with and without the magnitude bias defined in Eq.~(\ref{eq:Deltam}). The implications for key cosmological parameters are summarized in Table~\ref{t:no_mb}, along with the differences in best-fit $\chi^2$.

For both models, we find no preference for redshift-dependent bias from cosmological data. Specifically, we constrain the bias parameter to be $S_{\rm SN} = -0.005 \pm 0.003$ for $\Lambda$CDM and $0.007 \pm 0.008$ for $w_0w_a$ model. 
Given that the parameterization in~\cite{Son:2025rdz} exhibits a similar redshift dependence to our Eq.~\eqref{eq:Deltam}, we find that the value $S_{\rm SN}=0.03$ proposed there is disfavored by cosmological data.

Allowing for the bias has a minimal impact on cosmological parameters in $\Lambda$CDM, with practically no change in the mean values or uncertainties of the parameters. 

There is also no significant change in the nature of dark energy dynamics inferred from fitting the $w_0w_a$CDM model. The uncertainties in $w_0$ and $w_a$ increase, as expected from adding a new parameter, and there is a minor shift in the mean values towards a larger $w_0$ and smaller $w_a$. This shift yields a smaller value of the Hubble constant, $H_0 = 66.3 \pm 1.5\,{\rm km/s/Mpc}$, compared to $H_0 = 67.54 \pm 0.58\,{\rm km/s/Mpc}$ obtained from the same data combination without the bias, although the Hubble tension is not necessarily worse as the uncertainty in $H_0$ is notably increased. To see the reason for a smaller $H_0$, first note that the distance to recombination 
\begin{equation}\label{eq:DA}
    D_A = \frac{1}{H_0}\int_0^{z_*} \frac{dz'}{H(z')/H_0}
\end{equation}
is tightly constrained by the precise measurements of CMB acoustic peaks. A change in the late-time expansion history would alter $H(z')/H_0$ in the denominator, thus requiring a different value of $H_0$. In the DESI DR2 best-fit $w_0w_a$CDM model~\cite{DESI:2025zgx}, $w>-1$ at $z\lesssim 0.5$ and $w<-1$ at $z\gtrsim 0.5$. In such a model, the average dark energy density contributing to the integral through the denominator of (\ref{eq:DA}) is higher than its value today, requiring a lower $H_0$ compared to $\Lambda$CDM to keep the $D_A$ unchanged. Allowing for the magnitude bias yields a slightly larger $w_0$, further reducing the current dark energy density relative to its value at higher redshifts and thus requiring an even smaller Hubble constant.

\subsection{Implication of the redshift-dependent bias for the Hubble tension} \label{ss:sn_calib}

To explore the impact of allowing for the magnitude bias on the Hubble tension, we now fit the $\Lambda$CDM and the $w_0w_a$CDM models to the combination of CMB, BAO and the PP SNIa data with an additional prior on the intrinsic magnitude, $M_b = -19.253 \pm 0.027$, as measured by SH0ES~\cite{Riess:2021jrx}. Since $M_b$ is a parameter in the MCMC fit, the Hubble tension is assessed by comparing the posterior of $M_b$ to the SH0ES measurement. Our results are summarized in Table~\ref{t:with_mb} and Figs.~\ref{fig:cmb_bias} and \ref{fig:lumdist_combine}. 

Within $\Lambda$CDM, we find $S_{\rm SN}=-0.009\pm 0.003$, a $3\sigma$ deviation from zero. The corresponding posterior of the intrinsic magnitude, $M_b = -19.382 \pm 0.011$, is in $4.4\sigma$ tension with the SH0ES measurement, down from the $5.4\sigma$ tension when not allowing for the bias. The ability of a non-zero $S_{\rm SN}$ to relieve the Hubble tension within the $\Lambda$CDM model is limited because the redshift-dependent bias is exactly zero at $z=0$ by definition.

The ability of the redshift-dependent bias to relieve Hubble tension is enhanced when allowing for changes in the late-time expansion history. When fitting the $w_0w_a$CDM model, we find $S_{\rm SN}=-0.017\pm 0.004$, deviating from zero at $4.3\sigma$. We find $M_b = -19.308 \pm 0.023$, which reduces the Hubble tension to $1.5\sigma$, with the corresponding $H_0 = 71.30 \pm 0.85\,{\rm km/s/Mpc}$. The best-fit $\chi^2$ is improved by $-28.95$ compared to the $\Lambda$CDM model with no bias. Interestingly, the best-fit dynamical dark energy model can be well described with a constant $w<-1$. When fitting the $w$CDM model, we get similar results to $w_0w_a$CDM (see last column in Table~\ref{t:with_mb}), which means the $w_a$ degree of freedom is redundant. Unlike the best-fit $w_0w_a$CDM model discussed in the previous subsection, the average dark energy density that enters the denominator of Eq.~\eqref{eq:DA} is now lower than its value today, requiring a higher $H_0$ value to compensate for the difference in the expansion history. 

Fig.~\ref{fig:lumdist_combine} demonstrates how the Hubble tension is relieved. The figure shows the differences in the luminosity distance ($\Delta 5\log_{10}d_L(z)$) relative to the prediction of the PL-AP+DESI2+PP best-fit LCDM model. The BAO data points (only the $D_M/r_d$ or ``transverse'' BAO measurements), shown with green circles, are anchored to the sound horizon at recombination measured by CMB. The SNIa data points, shown with red triangles, are anchored to the SH0ES measurement of $M_b$. The difficulty in finding a cosmological model that can accommodate both datasets in Fig.~\ref{fig:lumdist_combine} is well-known as the essence of the Hubble tension ({\it e.g.} see also Fig.~4 in \cite{Pogosian:2021mcs} and Fig.~1 in \cite{Poulin:2024ken}). In our case, the tension is relieved thanks to the upward shift in the SNIa luminosities (shown with blue rectangles) at high redshifts, afforded by the negative magnitude bias, along with a downward trend in the best-fit-model $\log_{10}d_L$ at low redshifts, afforded by dynamical dark energy.

The high-$H_0$ solution (blue curve in Fig.~\ref{fig:lumdist_combine}) corresponds to a dynamical dark energy behavior distinct from that preferred by PL-AP+DESI2+PP without the correction (orange dash–dot curve in Fig.~\ref{fig:lumdist_combine}\footnote{When the SH0ES $M_b$ prior is not imposed, the red triangles are free to have a constant shift so that the orange dash-dot curve can fit them.}), and yields a worse but acceptable fit to the DESI BAO data (see Table~\ref{t:with_mb}). Consequently, given current cosmological data, this high-$H_0$ solution is favored only when the SH0ES $M_b$ prior is included, as illustrated in Fig.~\ref{fig:cmb_bias} (green {\it vs} orange). Its viability could change with future BAO measurements: stronger (weaker) evidence for the current DESI-preferred dynamical dark energy behavior (orange dash–dot) would tend to disfavor (favor) this high-$H_0$ solution.

It is interesting to ask if allowing for a non-zero curvature $\Omega_k$ could help with the Hubble tension in the presence of the redshift-dependent bias. As shown in Table~\ref{t:with_mb}, we find a minor preference for positive curvature in the $\Lambda$CDM model with an improvement the CMB fit and reduction of the Hubble tension to $4.0\sigma$. 
Varying $\Omega_k$ in the $w_0w_a$ model, on the other hand, makes practically no difference.

\section{Conclusions and Discussions}\label{sec:conclusion}

We investigate the impact of a redshift-dependent correction to SNIa magnitudes on the Hubble tension and the evidence for dynamical dark energy. This effect is parameterized by an amplitude that linearly relates the correction to the cosmic look-back time. We consider both the $\Lambda$CDM model and the $w_0w_a$ dynamical dark energy framework, with results summarized in Tables~\ref{t:no_mb} and \ref{t:with_mb}.
The redshift-dependent correction to SNIa magnitudes can also be interpreted as an ``effective" violation of the Distance Duality Relation~\cite{Tutusaus:2023cms,Teixeira:2025czm}.

For uncalibrated SNIa data (Table~\ref{t:no_mb}), we find no evidence for a nonzero redshift-dependent correction and no reduction in the Hubble tension. 
When incorporating SH0ES-calibrated SNIa data (Table~\ref{t:with_mb}), a nonzero correction is weakly favored within $\Lambda$CDM. 
More notably, in the presence of dynamical dark energy, a negative correction amplitude (corresponding to intrinsically dimmer SNIa at higher redshifts) is preferred at the $4.3\sigma$ level. This leads to a reconciliation between early- and late-Universe distance measurements, reducing the Hubble tension to $1.5\sigma$. The preferred model is well described by a constant equation of state, $w = -1.125 \pm 0.028$, with $H_0=71.42 \pm 0.68\,{\rm km/s/Mpc}$. If future BAO measurements show stronger evidence for the current DESI-preferred dynamical dark energy behavior, this high-$H_0$ solution will be disfavored. On the other hand, if the support for it persists with increasingly precise future measurements, the results presented in this work may motivate theoretical efforts to realize such a redshift-dependent correction to SNIa magnitudes.

\acknowledgments 
We thank Adam Riess for helpful comments. This research was enabled in part by support provided by the BC DRI Group and the Digital Research Alliance of Canada ({\tt alliancecan.ca}). SHM and LP are supported by the National Sciences and Engineering Research Council (NSERC) of Canada. M-X.L. is supported in part by a Canadian Institute for Theoretical Astrophysics (CITA) National Fellowship.

\FloatBarrier
\bibliographystyle{apsrev4-2}

%

\end{document}